\documentclass[aip,graphicx]{revtex4-1}
\usepackage[cp1251]{inputenc}
\usepackage[T2A]{fontenc}
\usepackage[english]{babel}
\usepackage{amssymb,latexsym,amsmath,amscd}
\usepackage{graphicx,color,framed}
\usepackage{xcolor}
\usepackage{siunitx} 
\begin{document}
%\selectlanguage{russian}
\selectlanguage{english}
\title{Nonlocal density functional theory of water taking into account many-body dipole correlations: Binodal and surface tension of 'liquid-vapour' interface}
\author{\firstname{Yu.A.} \surname{Budkov}}
\email[]{ybudkov@hse.ru}
%\homepage[]{Your web page}
%\thanks{}
%\altaffiliation{}
\affiliation{School of Applied Mathematics, Tikhonov Moscow Institute of Electronics and Mathematics, National Research University Higher School of Economics, Tallinskaya st. 34, 123458 Moscow, Russia}
\affiliation{G.A. Krestov Institute of Solution Chemistry of the Russian Academy of Sciences, Academicheskaya st., 1, 153045 Ivanovo, Russia}
\author{\firstname{A.L.} \surname{Kolesnikov}}
\email[]{kolesnikov@inc.uni-leipzig.de}
\affiliation{Institut f\"{u}r Nichtklassische Chemie e.V., Permoserstr. 15, 04318 Leipzig, Germany}
\begin{abstract}
In this paper we formulate a nonlocal density functional theory of inhomogeneous water. We model a water molecule as a couple of oppositely charged sites. The negatively charged sites interact with each other through the Lennard-Jones potential (steric and dispersion interactions), square-well potential (short-range specific interactions due to electron charge transfer), and Coulomb potential, whereas the positively charged sites interact with all types of sites by applying the Coulomb potential only. Taking into account the nonlocal packing effects via the fundamental measure theory (FMT), dispersion and specific interactions in the mean-field approximation, and electrostatic interactions at the many-body level through the random phase approximation, we describe the liquid-vapour interface. We demonstrate that our model without explicit account of the association of water molecules due to hydrogen bonding and with explicit account of the many-body electrostatic interactions at the many-body level is able to describe the liquid-vapour coexistence curve and the surface tension at the ambient pressures and temperatures. We obtain very good agreement with available in the literature MD simulation results for density profile of liquid-vapour interface at ambient state parameters. The formulated theory can be used as a theoretical background for describing of the capillary phenomena, occurring in micro- and mesoporous materials.     
\end{abstract}
\maketitle
\section{Introduction}
Theoretical description of water adsorption on micro- and mesoporous materials is a challenge problem for modern physical chemists and chemical engineers. Its great importance is due to numerous industrial applications and fundamental significance of describing water in a nanoconfinement. The examples of technological applications, where the description of water adsorption is highly relevant, are: characterizing of micro- and mesoporous materials at ambient conditions (i.e. determination of pore size distribution, surface area, accessible volume) \cite{Russo2007,Groenquist2019,Georgi2017,kolesnikov2018effects,kolesnikov2017pore,landers2013density}, modelling of water purification from toxic compounds (e.g., ions of heavy metals \cite{peng2017review}), description of mechanical stability of construction porous materials (concrete, wood, paper, etc.) during capillary condensation/evaporation cycles\cite{hamouda2002influence,jakovljevic2017influence,trong2018sorption,setzer1974surface} , {\sl etc}. All these examples have one common feature - the influence of inhomogeneity on the process.

Obtaining a correct description of inhomogeneous water requires a reliable theoretical model based on the first principles of statistical mechanics. Such a theoretical model must account for the electrostatic interactions (including short-range specific interactions, attributed to electron charge transfer) between water molecules and must be based on the nonlocal functional theory to describe correctly the liquid-vapour interface and temperature behavior of the surface tension. Despite the fact that up to now several theoretical models have been formulated \cite{Yang1994, Ding1987, Lischner2010, Jaqaman2004, Fu2005, Hughes2013, Krebs2014, Chuev2006, Trejos2019,trejos2018theoretical}, none of them satisfy the requirements formulated above.

In paper \cite{Yang1994} the authors formulated a density functional theory (DFT) taking account of the molecular structure of water within the TIP4P model \cite{Jorgensen1983} for describing the water liquid-vapour interface. The authors took into consideration the universal intermolecular interactions by the Lennard-Jones pair potential, whereas the electrostatic interactions between sites -- through the Coulomb potentials. To take into account for the electrostatic interactions in the total free energy functional, the authors used the mean-field approximation, expanding the anisotropic potential into the multipole series, truncated by the fifth order. They applied the local density approximation with the Weeks Chandler Andersen (WCA) procedure to the dispersion interactions \cite{andersen1971relationship}. Though the authors obtained rather satisfactory agreement with the experimental liquid-vapour coexistence curve (binodal), the values of the surface tension at all the temperatures were  highly overestimated. Such a discrepancy can be explained in two ways. Firstly, the authors used the local DFT and, thus, did not take into account the nonlocal packing effects which should be important for the water molecules, situated at the interface. Secondly, the authors did not consider the many-body electrostatic correlations of the water molecules which must be significant in condensed liquid phase. Indeed, while for a vapour phase the electrostatic correlations manifest themselves through the effective Keesom pairwise interactions, for the liquid condensed phase it is necessary to take into account the higher electrostatic correlations \cite{Budkov2018, Budkov2019}.  Nevertheless, from the result of the manuscript\cite{nezbeda2005towards}
one can conclude that the main electrostatic contribution comes from the short-range part of the Coulombic potential. The author shows that short-range attractive and repulsive interactions play the crucial role in the structure properties of polar and associating pure fluids. Later in the work \cite{rodgers2008interplay}
the authors pointed out that not accounting for the long-range electrostatic interactions leads to errors in the system with non-uniform geometries, however mostly errors arise in electrostatic properties.

%The importance of taking account of the electrostatic correlations at the many-body level was demonstrated in paper \cite{Alejandre1995}. The authors showed, that only accounting for the long-range electrostatic interactions between the sites of molecules in MD simulation within the SPC/E model by Ewald summation, one can achieve a good agreement between simulation and experimental results for the values of surface tension and densities of the coexisting phases at all temperatures. 

It is also necessary to mention the phenomenological density functional theories of water which do not take into account the electrostatic interactions explicitly \cite{Ding1987,Lischner2010,Jaqaman2004,Fu2005,Krebs2014,Hughes2013}. In paper \cite{Ding1987} the authors formulated a molecular DFT of water, which allowed authors to predict with good accuracy the  temperature of freezing at atmospheric pressure. To perform the numerical calculations, the authors used the experimental pair correlation functions oxygen-oxygen, oxygen-hydrogen, and hydrogen-hydrogen. Thus, despite the success of formulated molecular DFT, its application to different molecular systems requires the external experimental data regarding the site-site correlation functions (from X-ray scattering or full-atomistic computer simulations). In paper \cite{Lischner2010} the authors used a similar molecular phenomenological DFT approach to describe the liquid-vapour interface of water at a temperature of $298~K$. Despite the fact that the authors obtained a good fitting for experimental values of densities of the coexisting phases and surface tension, it remained unclear how this approach could describe these quantities at the other state parameters. Moreover, the formulated theory deals with the polynomial approximation for the excess free energy, whose phenomenological coefficients are not related to any statistical theory. In paper \cite{Jaqaman2004} the authors developed a phenomenological DFT which operates with the position-orientation number density of structured fluids. Despite the fact that this theory is based on the bulk equation of state, describing the anomalous behavior of water below $4^{0}~C$ and taking explicit account of the hydrogen bonding between the water molecules, whereas the free energy functional accounts for the nonlocality through the gradient term, the surface tension obtained is twice its experimental value. As the authors mentioned \cite{Jaqaman2004}, this discrepancy is most probably due to the simplicity of the gradient correction used. In paper \cite{Fu2005} a self-consistent DFT, taking explicit account of hydrogen bonding through the Statistical Associating Fluid Theory (SAFT) \cite{Wertheim1984}, is applied to investigating the phase behavior and surface tensions of water and aliphatic alcohols. The authors showed that for the bulk phases, their theory is reduced to an equation of state that provides an accurate description of saturation pressures as well as vapor-liquid phase diagrams. Near the critical region, the long-range fluctuations were taken into account using a renormalization group theory. It is worth mentioning the similar SAFT-based density functional approach \cite{Krebs2014,Hughes2013}, where the authors not only described the saturation pressure and surface tension, but also analyzed the effective interactions between the hydrophobic hard rods, immersed in liquid water. In the manuscript\cite{gloor2007prediction}, the authors used the SAFT-VR density functional theory in order to describe the vapor-liquid interface of associating and non-associating molecules, including water. The functional treats short-range repulsion, chain and association contributions in the local density approximation. A good description of both binodal and surface tension was achieved by including interfacial data in the optimization scheme \cite{gloor2007prediction}. In the work \cite{sundararaman2014efficient}, the authors developed classical density-functional theory of rigid-molecular fluid and applied it to calculations of thermodynamic and structural properties of water. The used functional contains the hard-spherical contribution (White Bear mark II), contribution of the attractive Van der Waals interactions (on the weighted-density approximation level), and electrostatic contribution, described within the mean-field approximation. The using hard-sphere diameter as a fitting parameter allowed authors to describe surface tension with good accuracy.

Despite the fact that in the mentioned papers the authors obtained a very good fitting for the saturation pressure and liquid-vapour surface tension, these are only a few theories accounting explicitly the electrostatic interactions between water molecules. The existing SAFT-based density functional theories can be considered as good tools for chemical engineering applications, but cannot answer the physically reasonable question: How big the contribution of the electrostatic interactions between water molecules to the surface tension of the liquid-vapour interface? Moreover, in principle, the SAFT-based theories in their present form cannot elucidate the role of the electrostatic correlations in the capillary phenomena taking place with confined water. 

One of the simplest water models is the SPC/E  - a 3-site model. Each site carries a point-like charge and additionally, the Lennard-Jones interaction potential between oxygen atoms is applied. Despite the fact that SPC/E model does not take into account hydrogen bonding explicitly, it can reproduce some thermodynamic and structural properties with sufficient accuracy \cite{Alejandre1995, vega2007surface}. Inspiring the fact that such a simple model allowed the authors to describe equilibrium water in MD simulations, we will formulate a nonlocal simple DFT approach for the description of inhomogeneous liquid water. Describing water molecules as spherically symmetric dipolar particles, having two oppositely charged sites and taking into account the many-body electrostatic correlations, universal inter molecular interactions, and short-range specific interactions, related to electron charge transfer, we will describe with good accuracy the binodal and the surface tension at ambient conditions. In the framework of the formulated theory, we will elucidate the role of different inter molecular interactions in the chemical potential of water at the liquid-vapour interface.

\section{Theory of bulk water}
Let us formulate a simple statistical theory of liquid (vapour) water in the bulk phase. We describe each water molecule as two sites with charges $\pm q$, separated by distance $l$, so that the dipole moment is $p=ql=1.85~D$.  We assume that sites with charges $-q<0$ interact with each other through the Coulomb potential, the Lennard-Jones (LJ) potential, and the attractive square-well short-range potential (see the description below). The sites with a charge $+q>0$ interact with all types of sites through the Coulomb potential only. We would like to note that the LJ potential describes the excluded volume and dispersion interactions, while the square-well potential takes into account at the primitive level the short-range specific interactions between water molecules. In other words, in this study we do not take explicit account of the association between water molecules, resulting in the formation of a hydrogen bond network. Instead, we replace the asymmetric chemical interactions by the effective spherically symmetric short-range square-well potential. Such a simplification is quite acceptable, because in this study we do not consider the fine effects, related to the network of the hydrogen bonds (see, for instance, \cite{Jorgensen1998,Holten2012}). As we will show below, such a primitive description of the effect of chemical interactions will allow us to describe the liquid-vapour coexistence curve and surface tension of water at the normal conditions that are the goal of the present study. Note that explicit account of the association effect was made in the general context in papers \cite{SEGURA1997,Yu2002,Haghmoradi2016,Erukhimovich2002,Trejos2019,trejos2018theoretical}. We also neglect the effect of static electronic polarizability of water molecules, as it is much less than the orientation polarizability, related to the permanent dipole moment at the normal temperature. Due to the fact that our model is the first statistical theory of liquid water taking into account dipole correlations of molecules at the many-body level within the random phase approximation, we have neglected for simplicity the orientation nonlinear effects, such as formation of chain-like clusters in accordance with $"$head-to-tail$"$ mechanism taking place usually in magnetic and ferroelectric fluids \cite{klapp2005dipolar,budkov2019nonlocal}. As we will show below, such an assumption allows us to describe successfully the phase coexistence curve and surface tension of liquid-vapour interface of water at the ambient conditions.

Bearing in mind all the above written model assumptions, we can write the density of Helmholtz free energy of water as follows:
\begin{equation}
f(\rho,T) = f_{id}(\rho,T)+f_{ex}(\rho,T),
\end{equation}
where 
\begin{equation}
f_{id}(\rho,T)=\rho k_{B}T\left(\ln(\rho \lambda^3)-1\right)
\end{equation}
is the free energy of the ideal gas.
The excess free energy can be written in the following form
\begin{equation}
f_{ex}(\rho,T)=f_{sr}(\rho,T) + f_{el}(\rho,T),
\end{equation}
where
\begin{equation}
\label{fLJ}
f_{sr}(\rho,T)=\rho k_B T \left(-\ln(1 - \eta) + \frac{3\eta}{1 - \eta} + \frac{3 \eta^2}{2(1 - \eta)^2}\right)+\frac{1}{2}B \rho^2,
\end{equation}
is the contribution of the short-range interactions, including the excluded volume interactions, dispersion interactions, and short-range specific interactions; $B$ is the parameter of attractive interactions accumulating the contributions from WCA attraction tail
\[
V_{WCA}(|\mathbf{r}|)=
\left\{
\begin{tabular}{ccc}
  $-\epsilon$ & ~if~ & $r < 2^{1/6}\sigma$ \\
  $4\epsilon\left[\left(\frac{\sigma}{r}\right)^{12} - \left(\frac{\sigma}{r}\right)^{6}\right]$ & ~if~ & $2^{1/6}\sigma < r < r_c$ \\
  $0$ & ~if~ & $r > r_c$ \\
  \end{tabular}
  \right\}
 \]
and 
\begin{equation}
V_{spc}(|\mathbf{r}|)=-\epsilon_{sw}\Theta(\sigma_{sw}/2-|\mathbf{r}|),
\end{equation}
where $r_{c}$ is the cutoff of LJ pairwise potential of interactions; $\Theta(r)$ is the Heviside step-function; $\sigma$ and $\epsilon$ are, respectively, the size and energy parameters of LJ potential; $\epsilon_{sw}$ and $\sigma_{sw}$ are, respectively, the energy and size parameters of the square-well potential. The latter describes the contribution of short-range specific interactions \cite{goodwin2017mean,budkov2018theory} which in our case is related to the electron charge transfer of the water molecules. We assume for our calculations that $r_{c}=5\sigma$. Such an assumption allows us to make our DFT calculations (see the next section) less time-consuming \cite{neimark1998pore}. The first term in (\ref{fLJ}) describes the contribution of the excluded volume interactions of the hard spheres within the Percus-Yewick approximation; $\eta= \pi d_{BH}^3\rho/6 $ is the packing fraction of hard spheres with the effective Barker-Henderson diameter, determined by the following Pade approximation \cite{Verlet1972a}
\begin{equation}
d_{BH} = \sigma\frac{1.068 \epsilon/k_B T + 0.3837}{\epsilon/k_B T + 0.4293}.
\end{equation}
 The second term in (\ref{fLJ}) describes the total contribution of the attractive interactions, thus $B$ can be defined as:
 \begin{equation}
     B =\int d\mathbf{r} \biggl(V_{WCA}(|\mathbf{r}|) + V_{spc}(|\mathbf{r}|) \biggr) = -\frac{32\sqrt{2}}{9}\pi\epsilon\sigma^3 + \frac{16}{3}\pi\epsilon\sigma^3\left[ \left(\frac{\sigma}{r_c}\right)^3 - \frac{1}{3}\left(\frac{\sigma}{r_c}\right)^9\right] -\frac{\pi\epsilon_{sw}\sigma_{sw}^3}{6}.  
 \end{equation}

The contribution of the electrostatic interactions, which in our case are reduced to the short-range dipole-dipole interactions, can be described by the free energy of the dipolar hard spheres (see Appendix 1):
\begin{equation}
f_{el} = -\frac{k_B T}{l^3} \left(1-\frac{3}{4}\alpha\right)\sigma(y),
\end{equation}
where $l$ is the dipole length and $y = \rho p^2/3 \epsilon_0 k_B T$, $p$ is the dipole moment. The auxiliary function
\begin{equation}
\sigma(y) = \frac{\sqrt{6}}{4\pi}\left[2(1+y)^{3/2}-2-3y\right]
\end{equation}
is also introduced; the compressibility factor $\alpha$ of the hard spheres (see Appendix 1) in Percus-Yevick approximation has the following form
\begin{equation}
\alpha=\frac{\eta(4-\eta)(2+\eta^2)}{(1+2\eta)^2}.
\end{equation}
 Before that, we talk only about the thermodynamic properties, however, it is interesting to study also the structural ones. Here, we compare calculated structure factor and pair correlation function of the homogeneous water at 298 K with presented literature experimental data and DFT calculations. In general, there are two methods to calculate pair correlation function within DFT - test particle method and method base on the Ornstein-Zernike (OZ) equation. Recently \cite{archer2017standard}, it was shown that the first one gives more accurate results and especially enforce the correct behavior at small distances. However, we unable to use it due to the lack of explicit known form of the effective electrostatic pair potential, thus we will use OZ-based method (see  Appendix IV). The Fourier transform of structure factor in the homogeneous limit can be calculated by the standard expression \cite{hansen1990theory}:
\begin{equation}
S(k) = \frac{1}{1 - \rho c^{(2)}(k)},
\end{equation}
where $k=|\bold{k}|$; the pair correlation function can be expressed through structure factor as:
\begin{equation}
g(r) = 1 + \frac{1}{2 \pi^2 r \rho } \int \limits_0^{\infty} dk k\sin(kr)(S(k) - 1). \label{g_eq}
\end{equation}

\begin{figure}[h!]
\center{\includegraphics[width=1\linewidth]{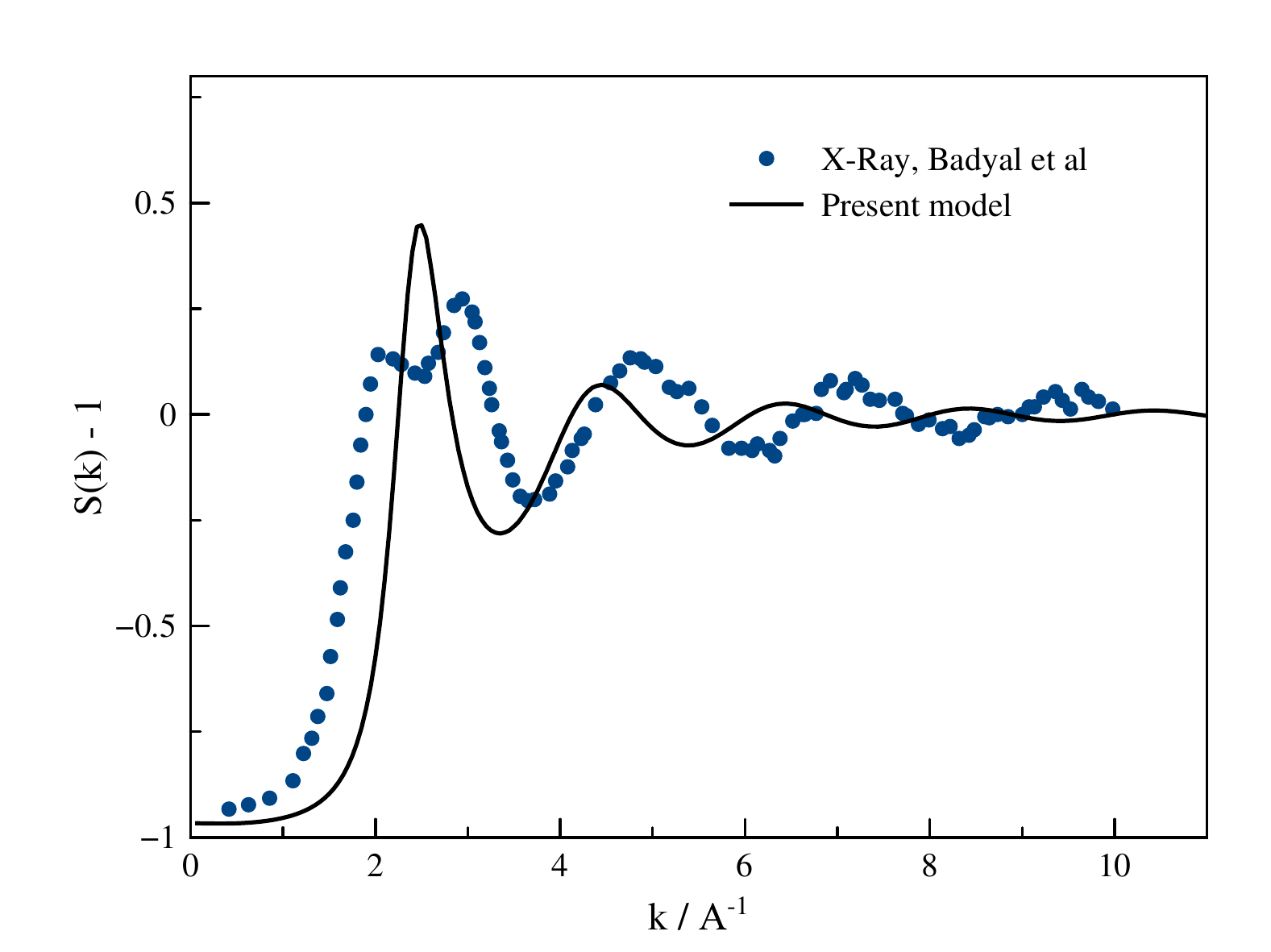}}
\caption{Comparison between calculated structure factor at ambient temperature and average intermolecular structure factor obtained from x-ray measurements\cite{badyal2000electron}. }
\label{struct}
\end{figure}

\begin{figure}[h!]
\center{\includegraphics[width=1\linewidth]{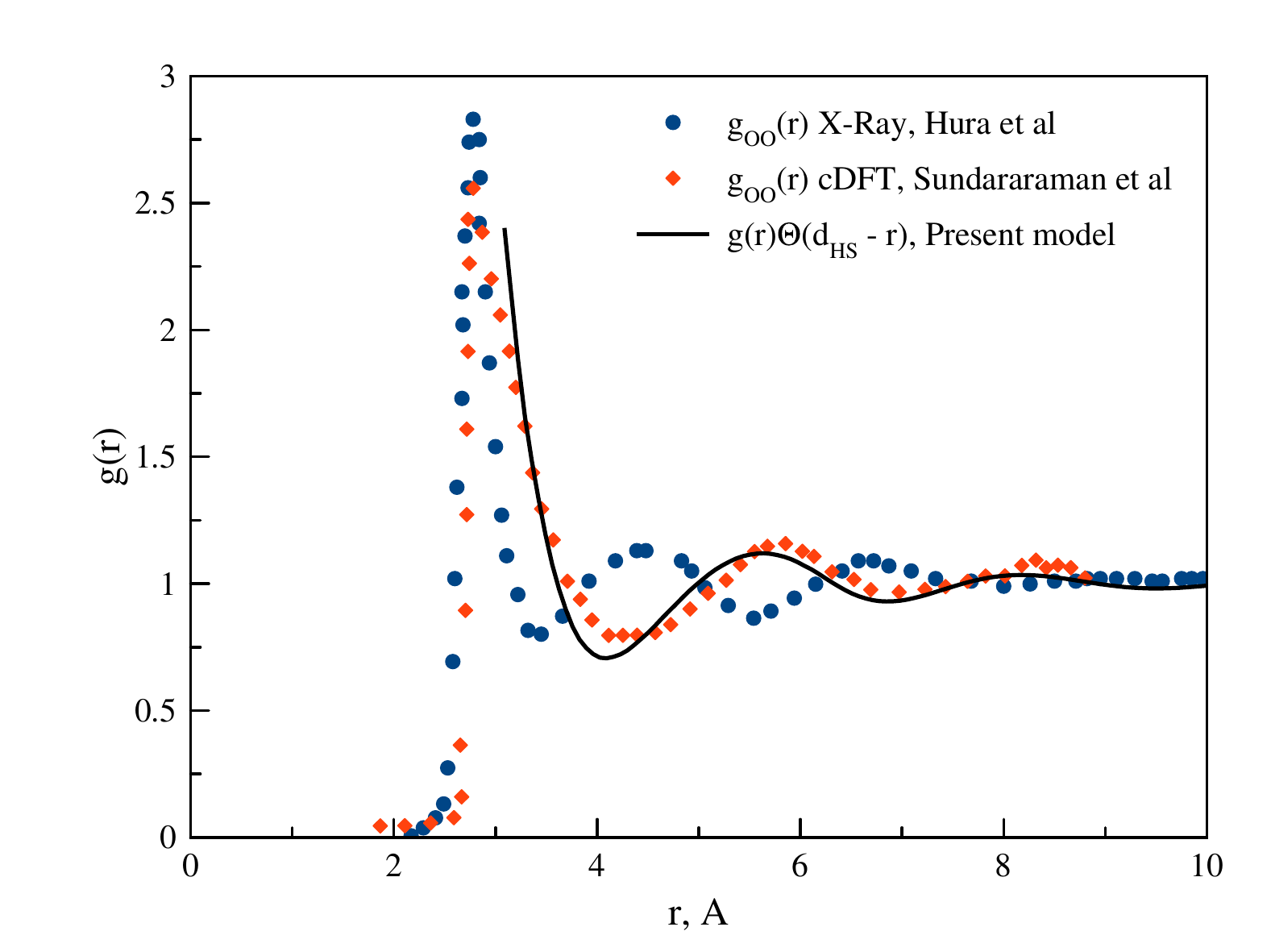}}
\caption{Comparison between pair correlation function calculated within the present approach, spherically averaged site-site correlation function oxygen-oxygen for scaler-EOS water functional\cite{sundararaman2014efficient} and x-ray measurements at ambient conditions.\cite{hura2000high}  }
\label{gOO}
\end{figure}

Figures \ref{struct} and \ref{gOO} represent the comparison between structure factor calculated within this model, molecular density functional theory \cite{sundararaman2014efficient}, and X-Ray scattering experiments \cite{hura2000high,badyal2000electron}. We obtain only partial qualitative agreement between our model and X-Ray obtained structure factor. Our results does not demonstrate the split of first pick \cite{bopp1996static} and the next picks, on bigger $k$, are shifted to the left. The first pick approximately corresponds to the average position of the first experimental ones and the amplitudes of picks are in agreement with experimental data. By means of Eq.(\ref{g_eq}) we calculated the pair correlation function. The Fig. \ref{gOO} represents X-Ray obtained \cite{hura2000high} and calculated within cDFT \cite{sundararaman2014efficient} $g_{OO}$,  and $g(r)\Theta(r - d_{HS})$ calculated in the present study. We plotted only part of the whole pair correlation function (cutted approximately on the distance of effective hard-sphere diameter), due to the numerical artifacts on the low distances. The pair correlation function does not go to zero in the core region. Such behavior was already previously reported in the literature \cite{archer2017standard} and can be attributed to the accuracy of the used method. Nevertheless, the obtained pair correlation function is in a good agreement with the results of molecular density functional theory except the position of the first maximum. The latter is shifted to the higher values by approximately 0.3 $\si{\angstrom}$. However, we would like to note that both theories fail to reproduce the real structure of water and demonstrates a significant shift of the second and third picks.

\section{Nonlocal density functional theory of inhomogeneous water}
Based on the bulk theory which was formulated in the previous section, let us formulate a density functional theory of inhomogeneous water. We start from the grand thermodynamic potential of inhomogeneous water in external potential field with the potential energy $V_{ext}(\bold{r})$, which can be written in the following form
\begin{equation}
\Omega[\rho(\mathbf{r})] = F_{id}[\rho(\mathbf{r})] + F_{ex}[\rho(\mathbf{r})]+\int_V d\mathbf{r} \rho(\mathbf{r}) V_{ext}(\mathbf{r}) - \mu\int_V d\mathbf{r} \rho(\mathbf{r}), \label{GTP}
\end{equation}
where $V$ is system volume, $F_{id}[\rho(\mathbf{r})]$ is the free energy of the ideal gas, $F_{ex}[\rho(\mathbf{r})]$ is the excess free energy of water, $\mu$ is the chemical potential and $\rho(\bold{r})$ is the single-particle density. Note that within this consideration $\rho(\bold{r})$ is the average density of the centers of mass of the water molecules. Thus, we construct the effective nonlocal DFT in terms of the simple fluid theory \cite{hansen1990theory} for water that is a molecular liquid in its nature. Such a simplification can be considered as an example of coarse-graining. The ideal gas free energy is
\begin{equation}
F_{id}[\rho(\mathbf{r})] = k_B T \int_V d\mathbf{r} \rho(\mathbf{r})[\ln(\lambda^3\rho(\mathbf{r})) - 1]  
\end{equation}
and the excess free energy, in turn, consists of several parts:
\begin{equation}
F_{ex}[\rho(\mathbf{r})] = F_{sr}[\rho(\mathbf{r})] + F_{el}[\rho(\mathbf{r})],
\end{equation}
where
\begin{equation}
F_{sr}[\rho(\mathbf{r})] = F_{fmt}[\rho(\mathbf{r})] + F_{att}[\rho(\mathbf{r})]
\end{equation}
with the Helmholtz free energy $F_{fmt}[\rho(\mathbf{r})] $ of hard spheres with the effective BH diameter $d_{BH}$ within the fundamental measure theory (FMT) \cite{rosenfeld1989free}, namely
\begin{equation}
F_{fmt}[\rho(\mathbf{r})] = \int_V d\mathbf{r} \Phi(\{n\}),
\end{equation}
where the free energy density is
\begin{equation}
\Phi(\{n\}) =k_{B}T\left(-n_0\ln(1 - n_3) + \frac{n_1 n_2 - \mathbf{n}^{(v)}_1 \mathbf{n}^{(v)}_2}{1 - n_3} + \frac{n_2^3 - 3 n_2 \mathbf{n}^{(v)}_2 \mathbf{n}^{(v)}_2}{24 \pi (1 - n_3)^2}\right),
\end{equation}
which depends on six weighted densities:  
\begin{equation}
n_{\alpha}(\mathbf{r}) = \int_V d\mathbf{r}^{\prime}\omega_{\alpha}(\mathbf{r} - \mathbf{r}^{\prime}) \rho(\mathbf{r}^{\prime}),
\end{equation}
where $\omega_{\alpha}(|\mathbf{r} - \mathbf{r}^{\prime}|)$ are the weight functions. Four of them are scalar values: $\omega_{2}(\mathbf{r}_{12}) = \delta(d_{BH}/2 - |\mathbf{r}_{12}|)$, $\omega_{3}(\mathbf{r}_{12}) = \Theta(d_{BH}/2 - |\mathbf{r}_{12}|)$, $\omega_{1}(\mathbf{r}_{12}) = \omega_{2}(\mathbf{r}_{12}) / (2\pi d_{BH})$, $\omega_{0}(\mathbf{r}_{12}) = \omega_{2}(\mathbf{r}_{12}) / (\pi d_{BH}^2)$,  and two are vector values: $\mathbf{\omega}^{(V)}_{2}(\mathbf{r}_{12}) = \mathbf{r}_{12}/r_{12}\delta(d_{BH}/2 - |\mathbf{r}_{12}|)$ and $\mathbf{\omega}^{(v)}_{1}(\mathbf{r}_{12}) = \mathbf{\omega}^{(v)}_{2}(\mathbf{r}_{12})/(2\pi d_{BH})$. The functional $F_{att}[\rho(\mathbf{r})]$ is the contribution from the given above attractive potentials $V_{WCA}(\bold{r})$ and $V_{spc}(\bold{r})$:
\begin{equation}
F_{att}[\rho(\mathbf{r})] = \frac{1}{2}\int\int_V d\mathbf{r}_1d\mathbf{r}_2\rho(\mathbf{r}_1)\rho(\mathbf{r}_2)\biggl[V_{WCA}(|\mathbf{r}_1 - \mathbf{r}_2|) + V_{spc}(|\mathbf{r}_1 - \mathbf{r}_2|)\biggr].
\end{equation}

Note that we do not know the real electrostatic free energy functional of dipolar hard spheres. On the other hand, due to the fact that thermodynamic properties of strongly inhomogeneous confined polar fluids must be very different from those are in the bulk phase, we cannot use the local density approximation for the electrostatic free energy functional. Nevertheless, since we know the approximate expression for the electrostatic free energy of the dipolar hard spheres system for the bulk phase, we can construct the phenomenological weighted density functional based on it, using the Curtin-Ashcroft-Tarazona approach \cite{curtin1985weighted,tarazona1985free,hansen1990theory}. We would like to note, that using of the weighted density functional approach can be justified by the fact that electrostatic interactions between water molecules in liquid phase manifest themselves as the effective short-range dipole-dipole interactions \footnote{This is related to the fact that the bulk electrostatic free energy of dipolar hard spheres system can be expanded into the power series on density (virial expansion), in contrast to the ionic systems, for which the electrostatic contribution to the total free energy is not an analytic function of density. As is well known, non-analytical behavior of electrostatic free energy of the ionic systems is determined by the long-range electrostatic interactions and translation entropy of ions \cite{landau2013course}. However, if the ionic groups form the electrically neutral clusters, as in dipolar \cite{Budkov2018,Budkov2019} or quadrupolar \cite{budkov2019statistical} fluids, the effective interactions between particles become short-range and, thereby, analyticity of the electrostatic free energy recovers.}. Thus, following the idea of the weighted density functional theory \cite{curtin1985weighted,tarazona1985free}, we treat the electrostatic free energy functional as follows:
\begin{equation}
F_{el}[\rho(\mathbf{r})] = \int_V d\bold{r}\rho(\mathbf{r}) \phi_{el}(\bar{\rho}(\mathbf{r})),
\end{equation}
where $\phi_{el}(\bar{\rho}(\mathbf{r})) = f_{el}(\bar{\rho})/\bar{\rho}$ is the free energy per fluid particle, depending on smoothed density, which, in turn, is determined in the following way:
\begin{equation}
\bar{\rho}(\mathbf{r}) = \int_V d\mathbf{r^{\prime}} \rho(\mathbf{r^{\prime}}) \omega_{el}(|\mathbf{r^{\prime}} - \mathbf{r}|),
\end{equation}
where $\omega_{el}(|\mathbf{r}|) = 3/(4\pi R_w^3)\Theta(R_w -|\mathbf{r}|)$ is the phenomenological weighted function with the phenomenological scale parameter $R_w$ determined the range of smoothing.

The equilibrium density profile is obtained from the minimization of the grand thermodynamic potential (\ref{GTP}), i.e. from the Euler-Lagrange equation
\begin{equation}
\frac{\delta\Omega[\rho(\mathbf{r})]}{\delta \rho(\mathbf{r})}=0
\end{equation}
or
\begin{equation}
\rho(\mathbf{r}) = \rho \exp \left[\beta(\mu_{ex} - V_{ext}(\mathbf{r})) + c^{(1)}(\bold{r})\right],
\end{equation}
where
\begin{equation}
\nonumber
c^{(1)}(\bold{r})=-\frac{\delta \left(\beta F_{ex}[\rho]\right)}{\delta\rho(\bold{r})}=c^{(1)}_{fmt}(\bold{r}) + c^{(1)}_{WCA}(\bold{r}) +c^{(1)}_{spc}(\bold{r}) + c^{(1)}_{el}(\bold{r}) 
\end{equation}
is the one-particle direct correlation function; $\beta=(k_{B}T)^{-1}$ is the inverse thermal energy; $\mu_{ex}=\mu_{ex}(\rho,T)$ is the bulk excess chemical potential, given in Appendix II. The contributions to the one-particle direct correlation function can be written as
\begin{equation}
c^{(1)}_{el}(\mathbf{r}) =-\frac{\delta \left(\beta F_{el}[\rho]\right)}{\delta\rho(\bold{r})}= -\beta\phi_{el}(\bar{\rho}(\mathbf{r})) - \beta\int_V d\mathbf{r^{\prime}} \rho(\mathbf{r^{\prime}}) \left[ \frac{\mu_{el}(\bar{\rho}(\mathbf{r^{\prime}}))}{\bar{\rho}(\mathbf{r^{\prime}})} - \frac{\phi_{el}(\bar{\rho}(\mathbf{r^{\prime}}))}{\bar{\rho}(\mathbf{r^{\prime}})} \right] \omega_{el}(|\mathbf{r} - \mathbf{r^{\prime}}|),
\end{equation}
\begin{equation}
c^{(1)}_{WCA}(\mathbf{r}) = -\frac{\delta \left(\beta F_{WCA}[\rho]\right)}{\delta\rho(\bold{r})}=-\beta \int_V d\mathbf{r^{\prime}}\rho(\mathbf{r^{\prime}})V_{WCA}(|\mathbf{r} - \mathbf{r^{\prime}}|),
\end{equation}
\begin{equation}
c^{(1)}_{spc}(\mathbf{r}) =-\frac{\delta \left(\beta F_{spc}[\rho]\right)}{\delta\rho(\bold{r})}= -\beta \int_V d\mathbf{r^{\prime}}\rho(\mathbf{r^{\prime}})V_{spc}(|\mathbf{r} - \mathbf{r^{\prime}}|),
\end{equation}
\begin{equation}
c^{(1)}_{fmt}(\mathbf{r}) =-\frac{\delta \left(\beta F_{fmt}[\rho]\right)}{\delta\rho(\bold{r})}= -\beta\sum\limits_{\alpha}\int_V d\mathbf{r^{\prime}}\frac{\partial \Phi(\{n\})}{\partial n_{\alpha}}\omega_{\alpha}(\mathbf{r^{\prime}} - \mathbf{r})
\end{equation}
Note that the formulated DFT for the bulk phase, where $V_{ext}(\bold{r})=0$ and, thus, $\rho(\bold{r})=\rho=const$, transforms into the bulk theory discussed in the previous section.

\section{Numerical results and discussions}
\begin{figure}[h!]
\center{\includegraphics[width=1\linewidth]{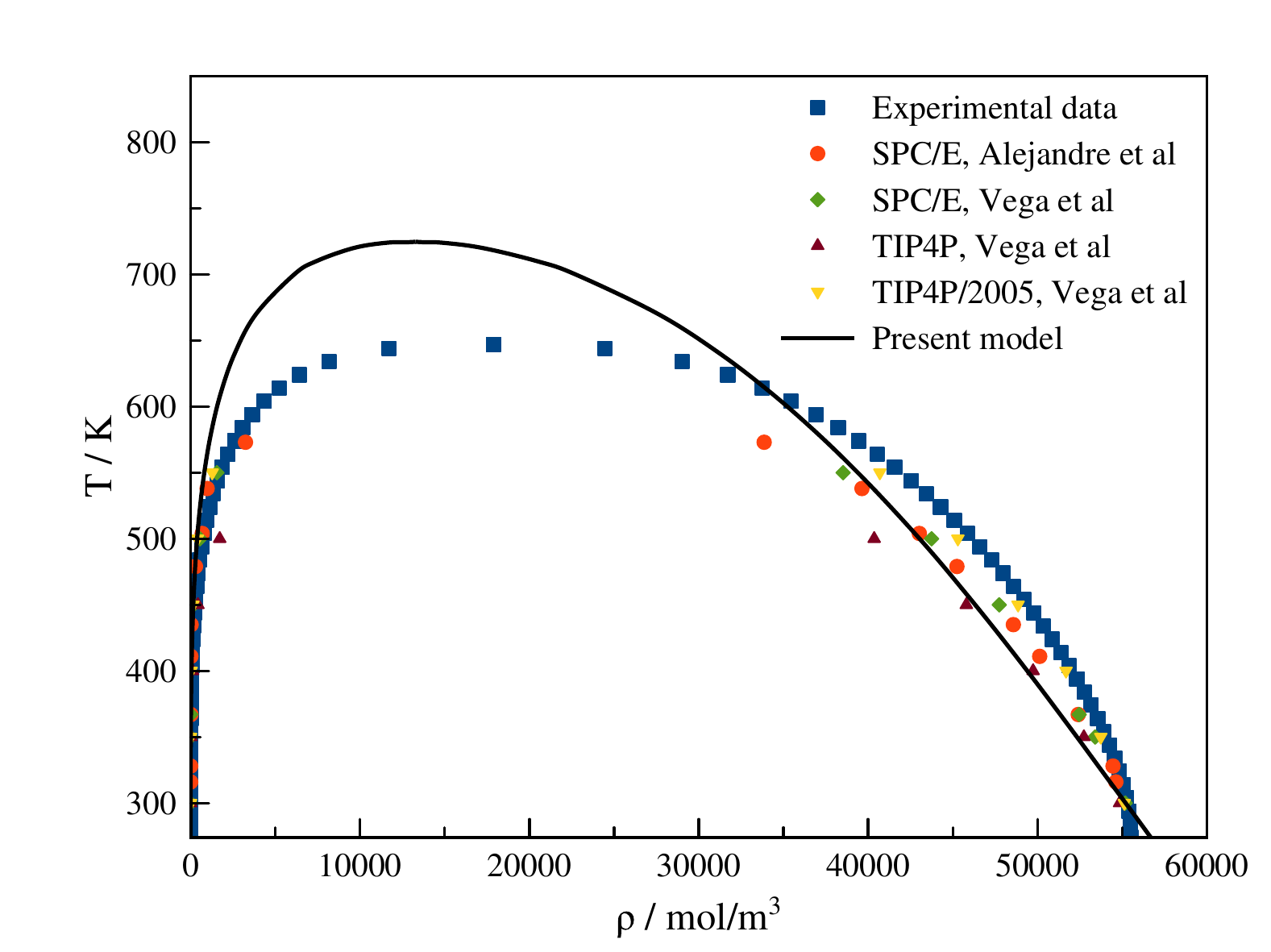}}
\caption{Liquid-vapour coexistence curves, obtained from the experiment (squares), molecular simulations (points), and present theoretical model (solid line).}
\label{binodal}
\end{figure}

Now we will consider the application of the model to the description of water liquid-vapour interface. We used the theory formulated above to fit the experimental values of coexisting densities and surface tension at the ambient temperature $T=298~K$. In order to calculate the densities of coexisting phases, we used the following system of equations, consisting of mechanical and chemical equilibrium conditions, i.e.
\begin{equation}
P(\rho_v,T) = P(\rho_l,T)~~ and~~ \mu(\rho_v,T) = \mu(\rho_l,T),
\end{equation}
where $\rho_v$ and $\rho_l$ are, respectively, the densities of the coexisting vapor and liquid phases. To calculate the equilibrium density profile, we used slit geometry with the length $H = 30\sigma - 60\sigma$ and discretization step along $z$ axis $\approx 0.02\sigma$. The surface tension was calculated by the following relation
\begin{equation}
\gamma=\Omega[\rho(z)]/A + PH,
\end{equation}
where $\Omega[\rho(z)]/A$ is the part of the grand thermodynamic potential per unit area $A$ corresponding to the liquid-vapour and $P$ is the pressure in the bulk liquid and vapour phases. The expressions for the pressure and chemical potential are given in Appendix II. The detailed description of DFT main equations in slit geometry is presented in Appendix III. Note that the obtained pressure at $T=298~K$ is deviated from the experimental value ($\approx 3141.7~kPa$) less than 1 $\%$. It is worth noting that the latent heat of vaporization at $298~K$ is $\approx 40.89~kJ/mol$, with experimental value $44~kJ/mol$. Thus, fitting the densities $\rho_v$ and $\rho_l$ and the surface tension $\gamma$ yields the following set of microscopic parameters:
\begin{center}
Table I: The obtained set of parameters for water.
\end{center}
\begin{center}
\begin{tabular}{ ||c c c c c c|| }
 $\sigma$ (\r{A}) & $\epsilon$ (kcal/mol) & $\sigma_{sw}$ (\r{A}) & $\epsilon_{sw}$(kcal/mol) & $l$ (\r{A}) & $R_w$ (\r{A})\\ 
 3.01 & 0.447 & 6.02 & 1.68 & 2.8 & 3.01 
\end{tabular}
\end{center}
Using these model microscopic parameters, we calculated the binodal and surface tension at other temperatures. Fig. \ref{binodal} shows the comparison between the binodal, calculated within our DFT, molecular simulations of Alejandre et al. \cite{Alejandre1995} and Vega et al\cite{vega2007surface}, and the experimental one \cite{lemmon1998thermophysical}.
\begin{figure}[h!]
\center{\includegraphics[width=1\linewidth]{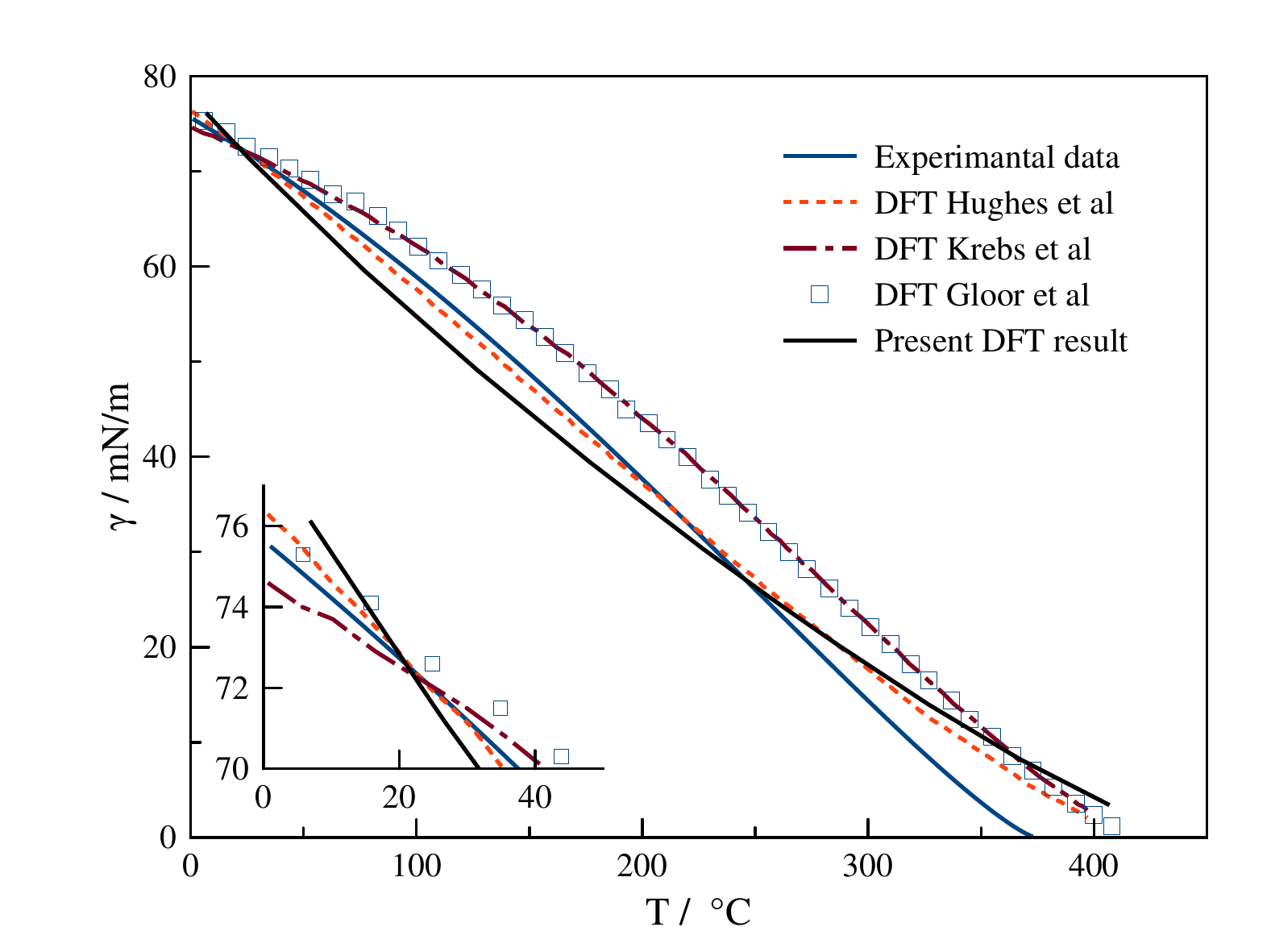}}
\caption{Dependences of surface tension on temperature, obtained from our density functional theory, the classical density functional theories presented in the literature and experiment. Worth noting that surface tension calculated within another two works \cite{sundararaman2014efficient,Yu2002} are in almost perfect agreement with exprimental data in the wide range of temperatures and will be barely visible on the present figure.}
\label{surf_tens}
\end{figure}

\begin{figure}[h!]
\center{\includegraphics[width=1\linewidth]{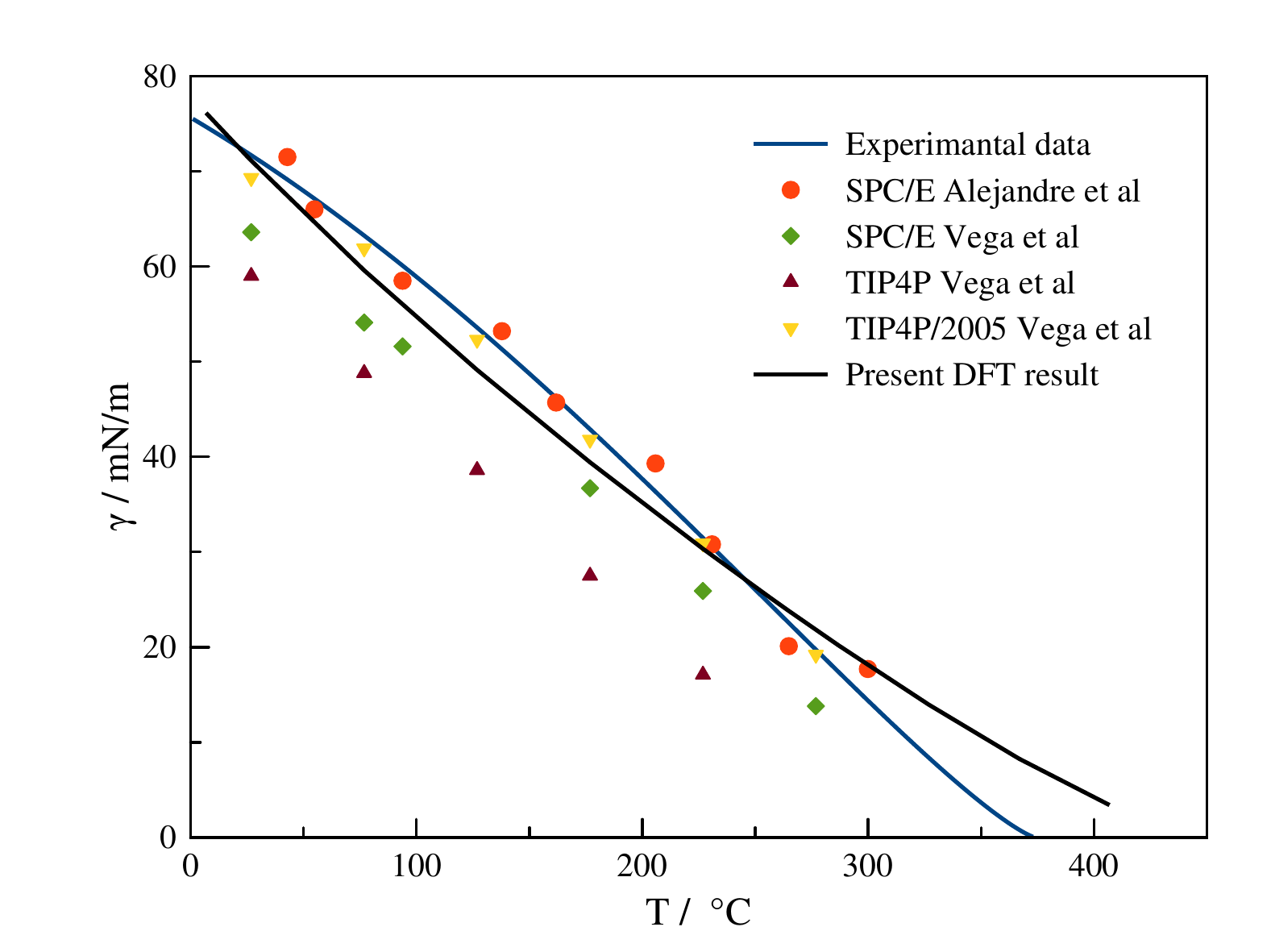}}
\caption{Dependences of surface tension on temperature, obtained from our density functional theory, molecular simulations, and experiment.}
\label{surf_tens_1}
\end{figure}
\begin{figure}[h!]
\center{\includegraphics[width=1\linewidth]{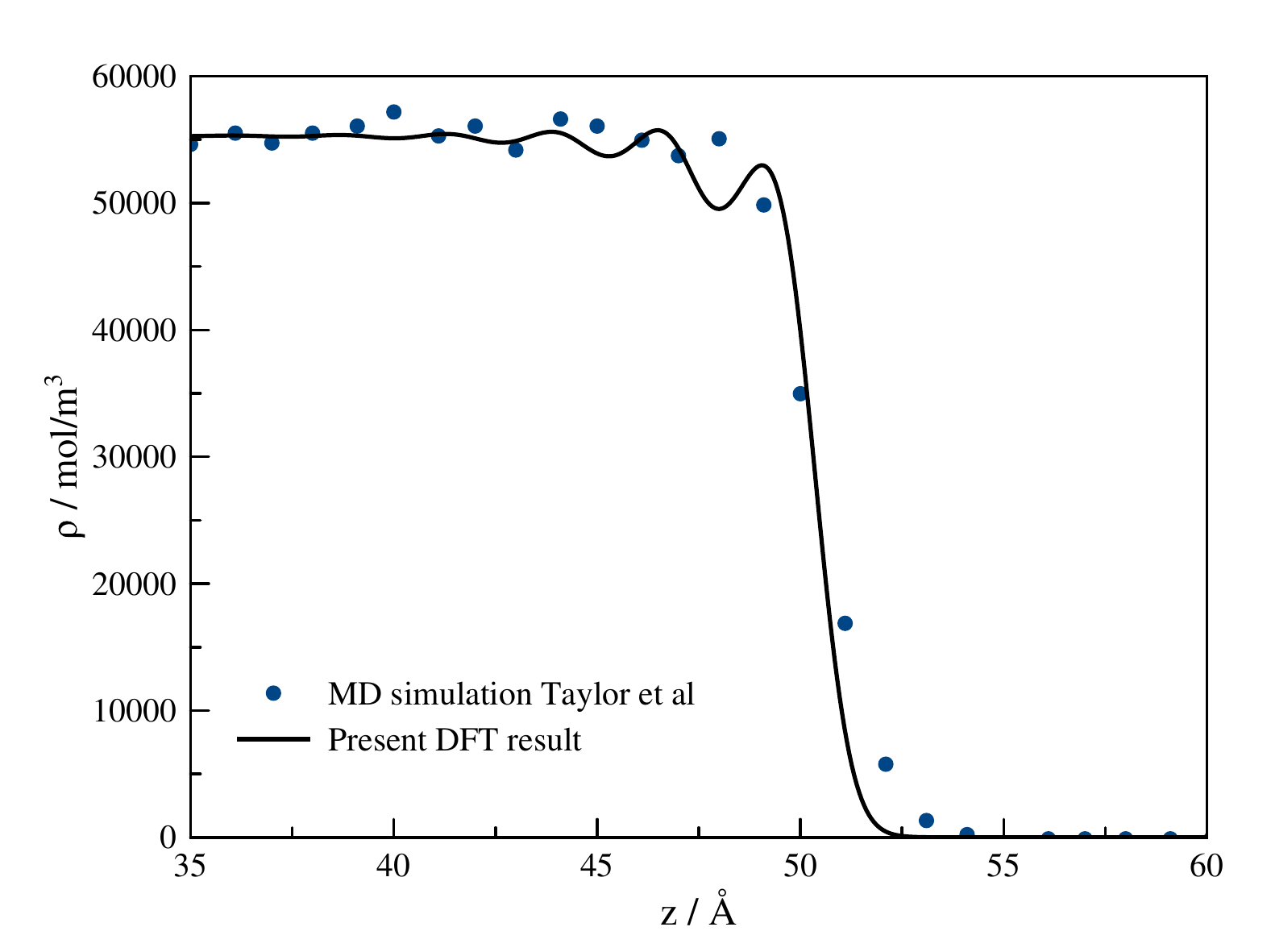}}
\caption{Density profile at the liquid-vapour interface of water at temperature $T=298~K$, calculated from our DFT (solid line) and from MD simulations of Taylor et al (symbols).}
\label{dens_prof1}
\end{figure}
\begin{figure}[h!]
\center{\includegraphics[width=1\linewidth]{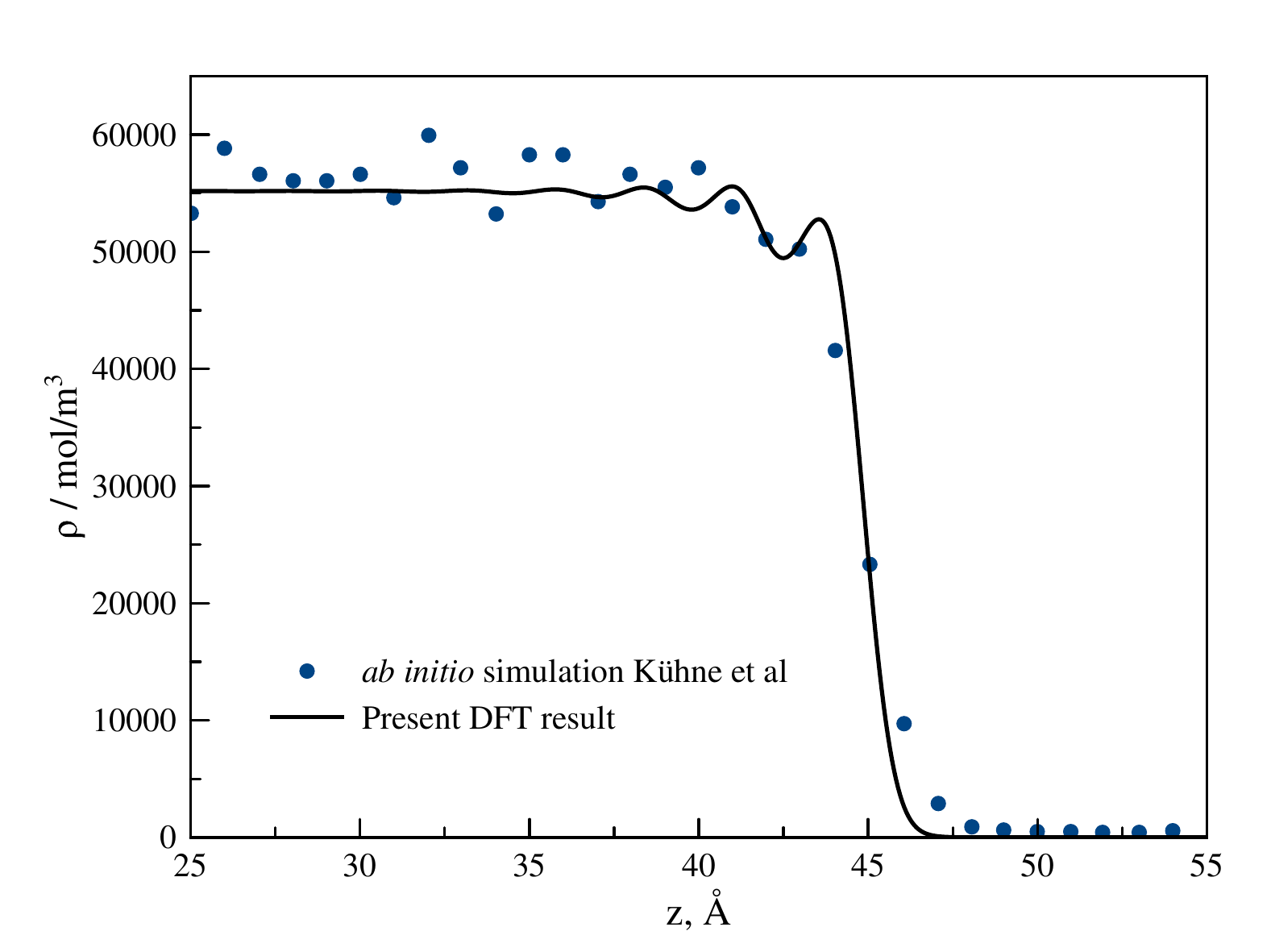}}
\caption{Density profile at the liquid-vapour interface of water at temperature $T=300~K$, calculated within our DFT (solid line) and taken from MD simulations of Kuhne et al (symbols).}
\label{dens_prof2}
\end{figure}

As one can see, the theoretical binodal is in very good agreement with both the MD simulation and the experiment at the ambient state parameters. However, our theory slightly underestimates the critical density and overestimates the critical temperature. However, this disagreement between the theoretical and experimental critical points is not very important for applications discussed in the Introduction. Fig. \ref{surf_tens} demonstrates a comparison between the dependences $\gamma~vs~T$, obtained from our DFT  and SAFT-based DFT \cite{Hughes2013,Krebs2014, gloor2007prediction}. The surface tension values calculated within our DFT agree very well with the experimental values at the ambient temperatures and deviate significantly only at sufficiently high temperatures that are close to the critical one. The latter is related to the fact that our theory cannot describe correctly the liquid-vapour equilibrium in the critical point vicinity. As is seen from fig. \ref{surf_tens}, the same picture is observed for the SAFT-based DFT \cite{Krebs2014,Hughes2013, vega2007surface}. We do not show the surface tension calculated in the works \cite{sundararaman2014efficient,Yu2002}. The results obtained in them are almost in a perfect agreement with experimental data.  Fig. \ref{surf_tens_1} demonstrates the comparison between our DFT, molecular simulations \cite{Alejandre1995, vega2007surface}, and experimental measurements \cite{lemmon1998thermophysical}. The surface tension obtained by different authors within one water model (SPC/E) are  different, which could be due to differences in the simulation procedures\cite{vinvs2016molecular}. The results of the present DFT is in quantitative agreement with TIP4P72005 water model. Further, we compared the density profile at the liquid-vapour interface calculated within our DFT the MD simulation with profiles available in the literature \cite{taylor1996molecular} and {\sl ab initio} Car-Parinello simulations \cite{kuhne2010new}. As Figures \ref{dens_prof1} and \ref{dens_prof2} show, our theory reproduces the simulation density profiles quite well. Note that DFT density profiles were shifted horizontally in order to match the regions of the steep descent of the simulated ones.

Despite the good consistent of the calculated surface tensions with experimental and simulation data, the density profiles are significantly steeper than simulation ones. That can be clearly identified by the comparison of the decaying length in the following equation \cite{hansen1990theory}:
\begin{equation}
\rho(z) = \frac{1}{2}(\rho_v + \rho_l) - \frac{1}{2}(\rho_l - \rho_v) \tanh\left(\frac{z-z_0}{d}\right),    
\end{equation}
where $\rho_v$ and $\rho_l$ are vapor and liquid densities, $z_0$ is a position of the Gibbs dividing surface and $d$ is the decay length. The same form was used in the work \cite{vega2007surface}, where the authors reported the value of "10-90" thicknesses ($t = 2.1972 d$) for density profiles calculated in the SPC/E, TIP4P and TIP4P/2005 water models at $300~K$. They are 3.39 $\si{\angstrom}$, 3.67 $\si{\angstrom}$ and 3.22 $\si{\angstrom}$, respectively. The value obtained in the present model is approximately $2$ times lower $\approx 1.76$ $\si{\angstrom}$.
Also, density profiles demonstrate slightly more pronounced oscillation behavior than those are obtained from molecular simulations.

\begin{figure}[h!]
\center{\includegraphics[width=1\linewidth]{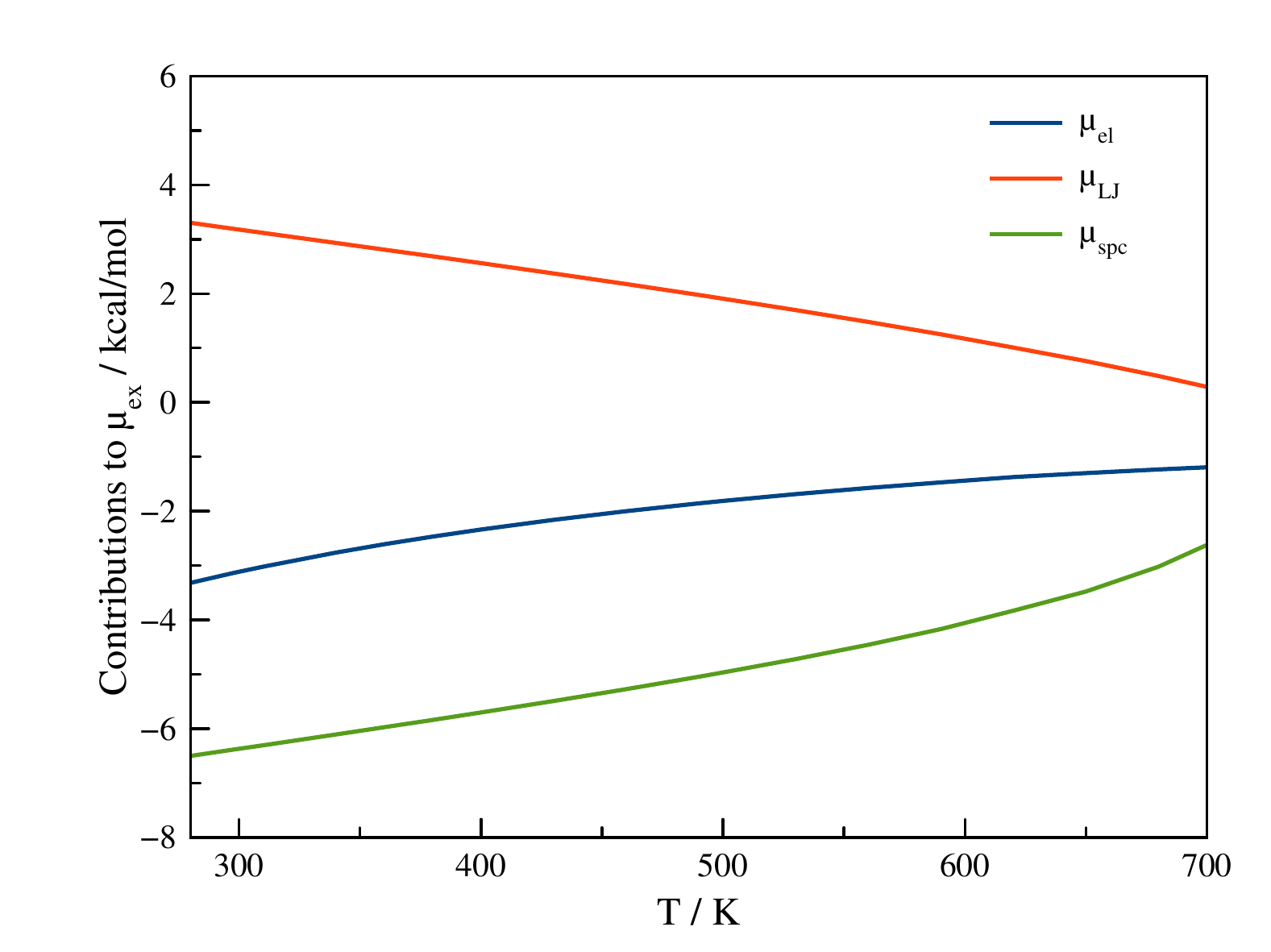}}
\caption{Contributions of different intermolecular interactions to excess chemical potential of water as functions of temperature, calculated along the liquid-phase branch of the binodal.}
\label{chem_pot}
\end{figure}

It is instructive to estimate the contributions of different intermolecular interactions to the excess chemical potential of liquid water predicted by our statistical model. Fig. \ref{chem_pot} shows these contributions as the functions of temperature, calculated along the liquid-phase branch $\rho_{l}=\rho_{l}(T)$ of the binodal. As is seen, the contribution of the universal intermolecular interactions ($\mu_{LJ}$) and electrostatic contribution ($\mu_{el}$) almost compensate for each other at all the temperatures, so that the total excess chemical potential in the liquid state region within our model is determined by the contribution of short-range specific interactions ($\mu_{spc}$). This prediction, in principle, could be verified by {\sl ab initio} Car-Parinello computer simulations \cite{kuhne2010new}.  
\section{Concluding remarks and prospects}
We have formulated a nonlocal density functional theory of inhomogeneous liquid water. We considered a water molecule as a couple of oppositely charged sites. The negatively charged sites interact with each other via the Lennard-Jones potential, square-well potential, and Coulomb potential, whereas the positively charged sites interact with all types of sites via the Coulomb potential only. Taking into account the nonlocal packing effects in the framework of the fundamental measure theory, dispersion and specific interactions in the mean-field approximation, and electrostatic interactions at the many-body level through the random phase approximation we have described the liquid-vapour interface. Namely, we have shown that our model without explicit account of the association of water molecules and with explicit account of the many-body electrostatic interactions at the many-body level is able to describe the liquid-vapour coexistence curve and the surface tension at ambient state parameters with good accuracy.

In conclusion, we would like to discuss the prospects of the formulated density functional theory. At first, this theory could be used as a theoretical background for describing the capillary phenomena, such as wetting/dewetting and capillary condensation/evaporation occurring at the solid surfaces of micro- and mesoporous materials. However, we cannot guarantee that application of this theory to the description of confined water will not change the values of the microscopic parameters of water molecules. However, we believe that the obtained microscopic parameters will be quite close to those obtained in the present study. The density functional theory of confined water will allow us to characterise micro- and mesoporous materials using the experimental adsorption isotherms of water vapour. Secondly, the formulated theory can be used for describing thermodynamic properties of other bulk and confined polar fluids, such as dimethylformamide, aliphatic alcohols, etc. However, these issues are the subject of forthcoming publications.  

\section*{Conflicts of interest}
There are no conflicts to declare.

\begin{acknowledgements}
The model development was supported by the RFBR according to research project No 18-31-20015. The project was partially supported by the RFBR according to research project No 18-29-06008. The authors thank Mikhail Kiselev for fruitful discussions and valuable comments.
\end{acknowledgements}

\section{Appendix I: Derivation of Helmholtz free energy of a system of dipolar hard spheres}
In this appendix we will briefly consider the fluctuation theory of complex fluids, whose electrically neutral molecules can be modelled as a set of clusters of spatially correlated charged centers with charges $q_{\alpha}$. A simplest example of such a fluid is a polar fluid, which we will consider in detail. More specifically, we will derive within the random phase approximation (RPA) an analytical expression for the Helmholtz free energy of the dipolar hard spheres system in the bulk.  We start from fluid partition function, which can be written as the following functional integral over the fluctuations of the local number densities $\rho_{\alpha}(\bold{r})$ of the charged centers as follows
\begin{equation}
Z=\mathcal{N}\int \prod\limits_{\alpha}^{}\mathcal{D}\rho_{\alpha}\exp\left[-\beta F_{0}[\{\rho_{\alpha}\}]-\beta U_{cl}[\{\rho_{\alpha}\}]\right],
\end{equation}
where $\beta=(k_{B}T)^{-1}$ is the inverse thermal energy, $F_{0}[\{\rho_{\alpha}\}]$ is the free energy functional of the reference system without Coulomb interactions between the particles; $\mathcal{N}$ is the normalization constant which will be specified below;
\begin{equation}
U_{cl}[\{\rho_{\alpha}\}]=\frac{1}{8\pi\varepsilon_{0}}\int d\bold{r}\int d\bold{r}^{\prime}\frac{\rho_{c}(\bold{r})\rho_{c}(\bold{r}^{\prime})}{|\bold{r}-\bold{r}^{\prime}|}
\end{equation}
is the energy of Coulomb interactions between the charged centers, $\rho_{c}(\bold{r})=\sum_{\alpha}q_{\alpha}\rho_{\alpha}(\bold{r})$ is the local charge density; $\varepsilon_{0}$ is the vacuum permittivity.

Further, we expand the free energy of the reference system into the functional power series near the average densities $\bar{\rho}_{\alpha}=\rho$ of the charged centers
\begin{equation}
F_{0}[\{\rho_{\alpha}\}]\approx F_{0}[\{\bar{\rho}_{\alpha}\}]+\frac{k_{B}T}{2}\int d\bold{r}\int d\bold{r}^{\prime}
\sum\limits_{\alpha,\gamma}G_{\alpha\gamma}^{-1}(\bold{r}-\bold{r}^{\prime})\delta\rho_{\alpha}(\bold{r})\delta\rho_{\gamma}(\bold{r}^{\prime}),
\end{equation}
where $\delta\rho_{\alpha}(\bold{r})=\rho_{\alpha}(\bold{r})-\bar{\rho}_{\alpha}(\bold{r})$ is the fluctuation of the local charge densities;
\begin{equation}
G^{-1}_{\alpha\gamma}(\bold{r}-\bold{r}^{\prime})=\frac{\delta^2}{\delta\rho_{\alpha}(\bold{r})\delta\rho_{\gamma}(\bold{r}^{\prime})}\left(\frac{F_{0}}{k_{B}T}\right)\biggr\rvert_{\rho=\bar{\rho}}
\end{equation}
is the inverse structure operator of the reference system for which we adopt the following approximation
\begin{equation}
G_{\alpha\gamma}^{-1}(\bold{r}-\bold{r}^{\prime})=W_{\alpha\gamma}^{-1}(\bold{r}-\bold{r}^{\prime})-c_{\alpha\gamma}(\bold{r}-\bold{r}^{\prime}),
\end{equation}
where $c_{\alpha\gamma}(\bold{r}-\bold{r}^{\prime})$ is the matrix of direct correlation functions of the sites that are not bonded to each other and 
\begin{equation}
W_{\alpha\gamma}(\bold{r}-\bold{r}^{\prime})=\rho\delta_{\alpha\gamma}\delta(\bold{r}-\bold{r}^{\prime})+\rho(1-\delta_{\alpha\gamma})g_{\alpha\gamma}(\bold{r}-\bold{r}^{\prime})
\end{equation}
is the matrix of the structure factors of molecules; $g_{\alpha\gamma}(\bold{r}-\bold{r}^{\prime})$ is the probability distribution function of the distance between the $\alpha^{th}$ and $\gamma^{th}$ sites. Thus, we have the following relation
\begin{equation}
Z\approx\exp\left[-\beta F_{0}[\{\bar{\rho}_{\alpha}\}]\right]\mathcal{N}\int \prod\limits_{\alpha}^{}\mathcal{D}\rho_{\alpha} \exp\left[-\frac{1}{2}\int d\bold{r}\int d\bold{r}^{\prime}
\sum\limits_{\alpha,\gamma}S_{\alpha\gamma}^{-1}(\bold{r}-\bold{r}^{\prime})\delta\rho_{\alpha}(\bold{r})\delta\rho_{\gamma}(\bold{r}^{\prime})\right],
\end{equation}
where 
\begin{equation}
S_{\alpha\gamma}^{-1}(\bold{r}-\bold{r}^{\prime})=G_{\alpha\gamma}^{-1}(\bold{r}-\bold{r}^{\prime})+\frac{q_{\alpha}q_{\gamma}}{4\pi\varepsilon_0 k_{B}T|\bold{r}-\bold{r}^{\prime}|}
\end{equation}
is the inverse structure operator in the random phase approximation. Further, choosing the normalization constant $\mathcal{N}$ from the condition that at $q_{\alpha}=0$ the electrostatic contribution to the free energy is equal to zero and calculating the Gaussian functional integral, we arrive at
\begin{equation}
F=F_{0}+F_{el},
\end{equation}
where 
\begin{equation}
\label{Fcorr3}
F_{el}=\frac{Vk_{B}T}{2}\int\frac{d\bold{k}}{(2\pi)^3}\left(\ln\left(1+\frac{\varkappa^2(\bold{k})}{k^2}\right)-\frac{\varkappa^2(\bold{k})}{k^2}\right)
\end{equation}
is the electrostatic free energy and
\begin{equation}
\varkappa^2(\bold{k})=\frac{1}{\varepsilon_0 k_{B}T}\sum\limits_{{\alpha\gamma}}q_{\alpha}q_{\gamma}G_{\alpha\gamma}(\bold{k})
\end{equation}
is the screening function. Note that we have subtracted from the final expression the electrostatic self-energy of the molecules
\begin{equation}
E_{self}=\frac{Vk_{B}T}{2}\int\frac{d\bold{k}}{(2\pi)^3}\frac{\varkappa^2({\bold{k}})}{k^2}.
\end{equation}
$G_{\alpha\gamma}(\bold{k})$ are the Fourier-images of the structure factors of the reference system which can be calculated from the following matrix relation \cite{borue1988statistical}
\begin{equation}
G^{-1}_{\alpha\gamma}(\bold{k})=W_{\alpha\gamma}^{-1}(\bold{k})-c_{\alpha\gamma}(\bold{k}).
\end{equation}
Now, we will turn to the theory of dipolar fluids, following from the general theory formulated above. In this case, we consider the dipolar particles as pairs of charges $\pm q$.  The relation for the structure factor of the dipolar particles in the RPA takes the following form
\begin{equation}
\label{eq:matrix1}
G^{-1}\left(\bold{k}\right)=W^{-1}\left(\bold{k}\right)-C(\bold{k})=
\begin{pmatrix}
\frac{1}{\rho\left(1-g^2(\bold{k})\right)}-c_{11}(\bold{k}) & -\frac{g(\bold{k})}{\rho\left(1-g^2(\bold{k})\right)}-c_{12}(\bold{k})\\
-\frac{g(\bold{k})}{\rho\left(1-g^2(\bold{k})\right)}-c_{12}(\bold{k}) & \frac{1}{\rho\left(1-g^2(\bold{k})\right)}-c_{22}(\bold{k})\\
\end{pmatrix}
,
\end{equation}
where
\begin{equation}
\label{eq:matrix1}
C\left(\bold{k}\right)=
\begin{pmatrix}
c_{11}(\bold{k}) & c_{12}(\bold{k})\\
c_{12}(\bold{k}) & c_{22}(\bold{k})\\
\end{pmatrix}
\end{equation}
is the matrix of the Fourier-images of the direct correlation functions of the reference system, whereas the structure matrix of dipolar molecules has the following form
\begin{equation}
\label{eq:matrix1}
W\left(\bold{k}\right)=
\begin{pmatrix}
\rho & \rho g(\bold{k})\\
\rho g(\bold{k}) & \rho\\
\end{pmatrix},
\end{equation}
where
\begin{equation}
g(\bold{k})=\int d\bold{r}g(\bold{r})e^{-i\bold{k}\bold{r}}
\end{equation}
is the characteristic function corresponding to the probability distribution function $g(\bold{r})$. The screening function \cite{Budkov2018,Budkov2019,budkov2019nonlocal,budkov2019statistical,borue1988statistical} is
\begin{equation}
\varkappa^2(\bold{k})=\frac{1}{\varepsilon_0 k_{B}T}\sum\limits_{{\alpha\gamma}}q_{\alpha}q_{\gamma}G_{\alpha\gamma}(\bold{k})=\frac{2\rho q^2}{\varepsilon_0 k_{B}T}\left(1-g(\bold{k})\right)Q(\bold{k}),
\end{equation}
where
\begin{equation}
Q(\bold{k})=\frac{1-\frac{\rho}{2}\left(c_{11}(\bold{k})+
c_{22}(\bold{k})+2c_{12}(\bold{k})\right)\left(1+g(\bold{k})\right)}{1-\rho\left(c_{11}(\bold{k})+
c_{22}(\bold{k})+2c_{12}(\bold{k})g(\bold{k})\right)+\rho^2\left(1-g^2(\bold{k})\right)\Delta(\bold{k})}
\end{equation}
and
\begin{equation}
\Delta(\bold{k})=c_{11}(\bold{k})c_{22}(\bold{k})-c_{12}^2(\bold{k}).
\end{equation}

Let us consider the reference system with $c_{11}(\bold{k})=c(\bold{k})$ and $c_{22}(\bold{k})=c_{12}(\bold{k})=0$, where $c(\bold{k})$ is the direct correlation function of the hard spheres. Thus, one of the sites (site $1$) is a center of a hard sphere, while the other site (site $2$) is a point-like one which does not correlate with the sites of all the species. Essentially, such a reference system describes a set of hard spheres with a grafted point-like particle which can freely penetrate inside the hard spheres. Thus, in this case we obtain
\begin{equation}
Q(\bold{k})=1+\frac{\rho}{2}h(\bold{k})\left(1-g(\bold{k})\right),
\end{equation}
where
\begin{equation}
h(\bold{k})=\frac{c(\bold{k})}{1-\rho c(\bold{k})}
\end{equation}
is the correlation function of the hard spheres \cite{hansen1990theory}.

Further, using the following model characteristic function
\begin{equation}
g(\bold{k})=\frac{1}{1+\frac{k^2l^2}{6}},
\end{equation}
the approximation
\begin{equation}
Q(\bold{k})\approx 1+\frac{\rho}{2}h(0)\left(1-g(\bold{k})\right)=1+\frac{Z_{0}-1}{2}\left(1-g(\bold{k})\right),
\end{equation}
and taking the integral (\ref{Fcorr3}), we arrive at the following relation for the density of the electrostatic free energy
\begin{equation}
\label{Fcorr4}
f_{el}=-\frac{k_{B}T}{l^3}\Lambda(y,\alpha),
\end{equation}
where $l$ is the dipole length and the auxiliary functions
\begin{equation}
\nonumber
\Lambda(y,\alpha)=\left(1-\frac{3\alpha}{4}\right)\sigma(y)-\frac{3\sqrt{6}\left((\alpha+4)(y^2+2y+y\sqrt{1+y})+8(1+\sqrt{1+y})\right)}{8\pi\left(1+\sqrt{1+y}\right)}
\end{equation}
\begin{equation}
\label{sigma2}
+\frac{3\sqrt{6}\exp\left[\frac{\alpha y}{4(1+\sqrt{1+y})^2}\right]\left(y\left(4+\sqrt{1+y}(\alpha+2)\right)+2y^2+4(1+\sqrt{1+y})\right)}{4\pi (1+\sqrt{1+y})}
\end{equation}
and
\begin{equation}
\sigma(y) = \frac{\sqrt{6}}{4\pi}\left[2(1+y)^{3/2}-2-3y\right]
\end{equation}
are introduced. Here, $\alpha=1-Z_{0}=1-\rho k_{B}T\chi_{0}$ with the isothermal compressibility $\chi_{0}$ of the hard spheres and $y=q^2l^2\rho/3\varepsilon_0 k_{B}T$. We would like to note that with good accuracy one can use the simplified relation for the electrostatic free energy of the dipolar hard spheres fluid
\begin{equation}
F_{el}=-\frac{Vk_{B}T}{l^3}\left(1-\frac{3}{4}\alpha\right)\sigma(y).
\end{equation}
In the Percus-Yewick approximation
\begin{equation}
\label{Fhs}
F_{0}=N k_B T\left(\ln(\lambda^3\rho)-1\right)+N k_B T \left(-\ln(1 - \eta) + \frac{3\eta}{1 - \eta} + \frac{3 \eta^2}{2(1 - \eta)^2}\right),
\end{equation}
taking into account the equation of state for the reference system
\begin{equation}
 P_{0}= \rho k_B T \frac{1 + \eta + \eta^2}{(1 - \eta)^3}
\end{equation}
and the relation for the compressibility $\chi_{0}=\rho^{-1}\partial{\rho}/\partial{P_{0}}$,
one can easily obtain a relation for the $"$compressibility factor$"$
\begin{equation}
\alpha=\frac{\eta(4-\eta)(2+\eta^2)}{(1+2\eta)^2},
\end{equation}
where $\eta=\pi d^3\rho /6$ is the packing fraction of the hard spheres; $d$ is the hard sphere diameter; $\lambda$ is the thermal de Broglie wavelength.

\section{Appendix II: Pressure and chemical potential in the bulk phase}
The total pressure $P$ and chemical potential $\mu$ are the sum of four contributions:
\begin{equation}
P = \rho\frac{\partial f}{\partial\rho}-f=P_{id}+P_{ex}=P_{id}+ P_{sr}+ P_{el}
\end{equation}
and 
\begin{equation}
\mu =\frac{\partial f}{\partial\rho}= \mu_{id}+\mu_{ex}=\mu_{id}+\mu_{sr}+\mu_{el}
\end{equation}
where $P_{id}=\rho k_{B}T$ and $\mu_{id}=k_{B}T\ln(\rho\lambda^3)$ are, respectively, the ideal gas pressure and chemical potential; $P_{ex}$ and $\mu_{ex}$ are the excess pressure and chemical potential, respectively, which, in turn, can be written through the sums of three contributions.

The ideal gas and short-range interaction contributions to the pressure and chemical potential are, respectively,
\begin{equation}
P_{id}+P_{sr}= \rho k_B T \frac{1 + \eta + \eta^2}{(1 - \eta)^3}+ \frac{1}{2}B\rho^2
\end{equation}
and
\begin{equation}
    \mu_{id}+\mu_{sr} = k_B T \ln(\rho\lambda^3) +  k_B T \left(-\ln(1 - \eta) +  \frac{\eta(14 - 13\eta + 5\eta^2)}{2(1 - \eta)^3} \right)+B\rho.
\end{equation}

The electrostatic contribution to the chemical potential takes the following form
\begin{equation}
\mu_{el} = -\frac{k_B T}{l^3}\left[\left(1 - \frac{3}{4}\alpha\right)\Theta_1 + \sigma(y)\Theta_2\right]   
\end{equation}
with the auxiliary functions
\begin{equation}
\Theta_1 =\frac{\sqrt{6}p^2}{4 \pi\epsilon_0 k_B T}(\sqrt{1 + y} - 1),
\end{equation}
\begin{equation}
\Theta_2 = -\frac{\pi d_{BH}^3}{2} \frac{(1-\eta)^3(2+\eta)}{(1+2\eta)^3}.
\end{equation}
The electrostatic contribution to the pressure can be calculated by the relation
\begin{equation}
P_{el} = \mu_{el}\rho - f_{el}.
\end{equation}

\section{Appendix III}
This Appendix presents detailed information about the contributions to the one-particle direct correlation function for the slit geometry. The direct correlation function of the specific interactions is
\begin{equation}
c^{(1)}_{spc}(z) =  \pi \beta  \epsilon_{sw}  \int\limits_{z - \sigma_{sw}/2}^{z + \sigma_{sw}/2} dz^{\prime} \rho(z^{\prime}) (\sigma_{sw}^2/4 - (z - z^{\prime})^2)
\end{equation}
and the contribution from the WCA potential $c^{(1)}_{WCA}(z)$
\begin{equation}
c^{(1)}_{WCA}(z) =  \int\limits_{z - r_c}^{z + r_c} dz^{\prime} \rho(z^{\prime})g_{WCA}(z,z^{\prime}),    
\end{equation}
where $g_{WCA}(z,z^{\prime})$ is
\[
g_{WCA}(z,z^{\prime})=
\left\{
  \begin{tabular}{ccc}
  $4\pi\epsilon\sigma^2 \psi(r_m) -\pi\epsilon(r_m^2 - (z^{\prime} - z)^2)$ & ~if~ & $|z - z^{\prime}|  < r_m$ \\
  $4\pi\epsilon\sigma^2 \psi(z^{\prime} - z)$ & ~if~ & $r_m<|z - z^{\prime}|< r_c$,\\
  $0$ & ~if~ & $|z - z^{\prime}|>r_c.$
  \end{tabular}
  \right\}
\]

Above we used the following notations $r_m = 2^{1/6}\sigma$ and $\psi(x)$ is
\begin{equation}
\psi(x) =  -\frac{1}{5}\left( \left(\frac{\sigma}{r_c}\right)^{10} - \left(\frac{\sigma}{x}\right)^{10}\right) + \frac{1}{2} \left( \left(\frac{\sigma}{r_c}\right)^{4} - \left(\frac{\sigma}{x}\right)^{4}\right).
\end{equation}

%\begin{align}
%g(z,z_0) &=  4\pi\epsilon\sigma^2\left[ -\frac{1}{5}\left( \left(\frac{\sigma}{r_c}\right)^{10} - \left(\frac{\sigma}{z-z_0}\right)^{10}\right) + \frac{1}{2} \left( \left(\frac{\sigma}{r_c}\right)^{4} - \left(\frac{\sigma}{z-z_0}\right)^{4}\right) \right]  \nonumber \\
%&\times\left[ H(z - (z_0 - r_c)) + H(z - (z_0 + r_m)) - H(z - (z_0 - r_m)) - H(z - (z_0 + r_c)) \right] \nonumber \\
%&+ \biggl( 4\pi\epsilon\sigma^2\left[-\frac{1}{5}\left( \left(\frac{\sigma}{r_c}\right)^{10} - \left(\frac{\sigma}{r_m}\right)^{10}\right) + \frac{1}{2} \left( \left(\frac{\sigma}{r_c}\right)^{4} - \left(\frac{\sigma}{r_m}\right)^{4}\right) \right]\nonumber \\
%&-\pi\epsilon(r_m^2 - (z - z_0)^2) \biggr)\times[H(z - (z_0 + r_m)) - H(z - (z_0 - r_m)]
%\end{align}
The three independent weighted functions are:
\begin{equation}
n_3(z) = \pi\int\limits_{z - R}^{z + R} dz^{\prime} \rho(z^{\prime}) (R^2 - (z - z^{\prime})^2),
\end{equation}
\begin{equation}
n_2(z) = 2\pi R \int\limits_{z - R}^{z + R} dz^{\prime} \rho(z^{\prime}),   
\end{equation}
\begin{equation}
\mathbf{n}^{(v)}_2(z) = 2 \pi \mathbf{e_z} \int\limits_{z - R}^{z + R} dz \rho(z^{\prime}) (z - z^{\prime}),  
\end{equation}
where $\mathbf{e_z}$ is the unit vector along the z-axis and $R = d_{BH}/2$.
Due to the fact that only three weighted functions are independent, we can express the hard spheres direct correlation function in the FMT approximation in the following way: 
\begin{equation}
c^{(1)}_{fmt}(z) =  2\pi\int\limits_{z - R}^{z+R} dz^{\prime}\left[\frac{1}{2}\frac{\partial \Phi}{\partial n_3}(R^2 - (z^{\prime} - z)^2) + R\frac{\partial \Phi}{\partial n_2} + \frac{\partial \Phi}{\partial n^{(v)}_2}(z^{\prime} - z) \right],    
\end{equation}
where the derivatives are determined as follows
\begin{equation}
\frac{\partial \Phi}{\partial n_3} = \frac{n_2}{4\pi R^2(1 - n_3)} + \frac{n_2^2 - (n^{(v)}_2)^2}{4\pi R (1 - n_3)^2} + \frac{n_2^3 - 3n_2 (n^{(v)}_2)^2}{12\pi(1-n_3)^3},
\end{equation}
\begin{equation}
\frac{\partial \Phi}{\partial n_2} = -\ln(1 - n_3)/(4\pi R) + \frac{n_2}{2\pi R(1 - n_3)} + \frac{n_2^2 - (n^{(v)}_2)^2}{8\pi(1-n_3)^2},
\end{equation}
\begin{equation}
\frac{\partial \Phi}{\partial n^{(v)}_2} = -\frac{n^{(v)}_2}{2\pi R(1 - n_3)} - \frac{n_2 n^{(v)}_2}{4\pi(1-n_3)^2}.
\end{equation}

The direct correlation function of electrostatic interactions is
\begin{equation}
c^{(1)}_{el}(z) =  -\beta\phi_{el}(\bar{\rho}(z)) - \frac{3\beta}{4 R_w^3}\int\limits_{-\infty}^{\infty} dz^{\prime} \rho(z^{\prime})  \left[ \frac{\mu_{el}(\bar{\rho}(z^{\prime}))}{\bar{\rho}(z^{\prime})} - \frac{\phi_{el}(\bar{\rho}(z^{\prime}))}{\bar{\rho}(z^{\prime})}\right] (R_w^2 - (z - z^{\prime})^2).
\end{equation}

\section{Appendix IV: direct correlation functions}
In this Appendix, we consider the calculations of the second direct correlation function $c^{(2)}(\bold{r},\bold{r}^{\prime})$ within the nonlocal density functional theory, considered in the main text. 
The second direct correlation function $c^{(2)}(\mathbf{r} ,\mathbf{r^{\prime}})$ is the sum of the following contributions:
\begin{eqnarray}
    c^{(2)}_{el} (\mathbf{r} ,\mathbf{r^{\prime}}) &=&-   \frac{2\beta(\mu_{el}(\bar{\rho}(\mathbf{r^{\prime}}))-\phi_{el}(\bar{\rho}(\mathbf{r^{\prime}})))}{\bar{\rho}(\mathbf{r^{\prime}})} \omega_{el}(|\mathbf{r} - \mathbf{r^{\prime}}|) \nonumber \\
    &-&\int_V d\mathbf{r^{\prime\prime}} \rho(\mathbf{r^{\prime\prime}}) \left[\frac{1}{\bar{\rho}( \mathbf{r^{\prime\prime}})}\frac{\partial \beta\mu_{el}(\bar{\rho}(\mathbf{r^{\prime \prime}}))}{\partial \bar{\rho}(\mathbf{r^{\prime\prime}})}  - \frac{2\beta(\mu_{el}(\bar{\rho}(\mathbf{r^{\prime \prime}})) - \phi_{el}(\bar{\rho}(\mathbf{r^{\prime \prime}}))}{\bar{\rho}^2(\mathbf{r^{\prime \prime}})} \right]\nonumber \\ &\times& \omega_{el}(|\mathbf{r} - \mathbf{r^{\prime\prime}}|) \omega_{el}(|\mathbf{r^{\prime}} - \mathbf{r^{\prime\prime}}|), \nonumber
\end{eqnarray}
\begin{equation}
    c^{(2)}_{WCA} (\mathbf{r} ,\mathbf{r^{\prime}}) = -\beta V_{WCA}(|\mathbf{r} - \mathbf{r^{\prime}}|),
\end{equation}
\begin{equation}
    c^{(2)}_{spc} (\mathbf{r} ,\mathbf{r^{\prime}}) = -\beta V_{spc}(|\mathbf{r} - \mathbf{r^{\prime}}|),
\end{equation}
\begin{equation}
    c^{(2)}_{fmt} (\mathbf{r} ,\mathbf{r^{\prime}}) = -\beta\sum\limits_{\alpha}\sum\limits_{\gamma} \int_V  d\mathbf{r^{\prime\prime}}\frac{\partial^2 \Phi(\{n\})}{\partial n_{\alpha}(\mathbf{r^{\prime\prime}})\partial n_{\gamma}(\mathbf{r^{\prime\prime}})}\omega_{\alpha}(\mathbf{r^{\prime\prime}} - \mathbf{r})\omega_{\gamma}(\mathbf{r^{\prime\prime}} - \mathbf{r^{\prime}}).
\end{equation}
%\begin{eqnarray}
% c^{(2)}(|\mathbf{r} - \mathbf{r^{\prime}}|) &=& c_{py}(|\mathbf{r} - \mathbf{r^{\prime}}|) - \beta V_{WCA} (|\mathbf{r} - \mathbf{r^{\prime}}|) - \beta V_{spc} (|\mathbf{r} - \mathbf{r^{\prime}}|)\\
     %&-&  \frac{2\beta(\mu_{el}(\rho)-\phi_{el}(\rho))}{\rho}\omega_{el}(|\mathbf{r} - \mathbf{r^{\prime}}|)\\ 
    %&-& \left[\frac{\partial \beta\mu_{el}(\rho)}{\partial \rho}  - \frac{2\beta(\mu_{el}(\rho) - \phi_{el}(\rho))}{\rho}\right] \int_V d\mathbf{r^{\prime\prime}}   \omega_{el}(|\mathbf{r} - \mathbf{r^{\prime\prime}}|) \omega_{el}(|\mathbf{r^{\prime}} - \mathbf{r^{\prime\prime}}|).
%\end{eqnarray}
The second direct correlation function for the homogeneous water in the Fourier-representation can be written as a sum of several contributions
\begin{equation}
\rho c^{(2)}(k) = \rho (c_{py}^{(2)}(k) +  c_{spc}(k) + c_{wca}(k) + c_{el}(k)),
\end{equation}
where
\begin{equation}
\rho c_{spc}(k) = 24 \eta \epsilon_{sw} \beta \left(\frac{r_{sw}}{d_{HS}}\right)^3 \frac{\sin(kr_{sw}) - kr_{rw}\cos(kr_{sw})}{(kr_{sw})^3},
\end{equation}
\begin{eqnarray}
\rho c_{WCA}(k) &=& 24 \eta \epsilon \beta \left(\frac{r_{m}}{d_{HS}}\right)^3 \frac{\sin(kr_{m}) - kr_{m}\cos(kr_{m})}{(kr_{m})^3} \nonumber \\
    &-& 96\beta\epsilon\eta\frac{\sigma}{k d_{HS}^3} \int\limits_{r_m}^{r_c} dr \sin(kr)\left[ \left(\frac{\sigma}{r}\right)^{11} - \left(\frac{\sigma}{r}\right)^5\right],
\end{eqnarray}
and
\begin{equation}
\rho c_{el}(k) = -\beta \rho \omega_{el}^2(k)\frac{\partial \mu_{el}}{\partial \rho} - 2\beta\rho\omega_{el}(k)(1 - \omega_{el}(k))\left(\frac{\mu_{el} - \phi_{el}}{\rho}\right).
\end{equation}
$c_{py}^{(2)}(k)$ is the Fourier image of the direct correlation function of the hard spheres system in the Perkus-Yevick approximation (see, for instance, \cite{hansen1990theory}), $r_{rw} = \sigma_{rw}/2$ and the Fourier-image of smoothing function $\omega_{el}$ is:
\begin{equation}
    \omega_{el}(k) = -3 \frac{kR_w \cos(kR_w) - \sin(kR_w)}{(kR_w)^3}.
\end{equation}

\bibliography{lit}
\end{document}